\documentclass[%
 aip,
 amsmath,amssymb,
 reprint,%
]{revtex4-1}

\usepackage{graphicx}
\usepackage{dcolumn}
\usepackage{bm}

\usepackage[utf8]{inputenc}
\usepackage[T1]{fontenc}
\usepackage{mathptmx}
\usepackage{etoolbox}

\usepackage{xcolor}

\makeatletter
\def\@email#1#2{%
 \endgroup
 \patchcmd{\titleblock@produce}
  {\frontmatter@RRAPformat}
  {\frontmatter@RRAPformat{\produce@RRAP{*#1\href{mailto:#2}{#2}}}\frontmatter@RRAPformat}
  {}{}
}%
\makeatother
\begin{document}

\preprint{AIP/123-QED}

\title[Nonlinear Transport in Carbon Quantum Dot Electronic Devices: Experiment and Theory]{Nonlinear Transport in Carbon Quantum Dot Electronic Devices: Experiment and Theory}
\author{Scott Copeland}
\affiliation{eM-TECH Inc.\ Framingham, MA, 01702 USA}
\author{Sungguen Ryu}
\affiliation{Institute for Cross-Disciplinary Physics and Complex Systems, IFISC (UIB-CSIC), Campus Universitat Illes Balears, E-07122 Palma, Spain
}
\author{Kazunari Imai}
\affiliation{NAMICS North America R\&D Center, DIEMAT Inc, Byfield, MA, US, 01922
}
\author{Nicholas Krasco}
\affiliation{NAMICS North America R\&D Center, DIEMAT Inc, Byfield, MA, US, 01922
}
\author{Zhixiang Lu}
\affiliation{NAMICS North America R\&D Center, DIEMAT Inc, Byfield, MA, US, 01922
}
\author{David S\'anchez}
\affiliation{Institute for Cross-Disciplinary Physics and Complex Systems, IFISC (UIB-CSIC), Campus Universitat Illes Balears, E-07122 Palma, Spain
}
\email{david.sanchez@uib.es}
\author{Paul Czubarow}
\affiliation{eM-TECH Inc.\ Framingham, MA, 01702 USA}

\date{\today}

\begin{abstract}
Carbon quantum dots (CQDs) are a promising material for electronic applications due to their easy fabrication and interesting semiconductor properties. Further, CQDs exhibit quantum confinement and charging effects, which may lead not only to improved performances but also to devices with novel functionalities. Here, we investigate the electronic transport of CQDs embedded on epoxy polymer. Our samples are coupled to interdigitated electrodes with individually addressable microelectrodes. Remarkably, the current-voltage characteristics show strongly nonlinear regimes at room temperature, ranging from Schottky diode to Coulomb blockade and even negative differential conductance behavior. We propose a master equation theoretical framework which allows us to compute current curves that agree well with the observations. This model emphasizes the importance of interacting dots and electron traps in generating a cohesive picture that encompasses all transport regimes. Overall, our results suggest that CQDs constitute a versatile materials platform for 3D integrated electronic purposes.
\end{abstract}

\maketitle

\textcolor{black}{As semiconducting conducting technologies develop, the need for 3D integrated solutions is of growing importance since electronic devices are reduced to increasingly smaller sizes. This requires new approaches to hardware to overcome tasks for device aspects that cannot be further scaled down. To this end, quantum dots are promising devices due to their high electric tunability, few-level spectrum and straightforward integration~\cite{kouwenhoven2001few}.
However, traditional semiconductor dots show unwanted drawbacks such as
intricate preparation involving imperfections~\cite{zwolak2024data}
and toxicity due to the release of metallic ions~\cite{larson2003water}.}

\textcolor{black}{In contrast, carbon quantum dots (CQDs)~\cite{lim2015carbon}
can be massively produced and are biocompatible for sensing and optical imaging~\cite{tian2020carbon,khan2023biocompatible}.
For electronic applications, CQDs can act as channels in field-effect transistors~\cite{kwon2013carbon} or form a basic material
in high-performance supercapacitors~\cite{permatasari2021carbon}.
Yet, a desired property toward 3D electronic integration that has thus far not been largely explored is surge suppression. This is commonly done with diode or varistor devices~\cite{he2019metal}. Ideally, these systems should exhibit high $\alpha$ values, where $\alpha$ is a coefficient that quantifies the nonlinearity strength of the current--voltage curve. Thus, high-$\alpha$ varistors allow for large current capacities and therefore serve as electrostatic protection elements.}

\textcolor{black}{Motivated by this quest of finding miniaturized structures that reliably protect externally exposed multichip modules~\cite{czubarow2019over}, the aim of this Letter is to report unique nonlinear electron transport characteristics of CQDs due to both their quantum confinement and large charging energies.}

\begin{figure}[t]
  \centering
  \includegraphics[width=0.45\textwidth,clip]{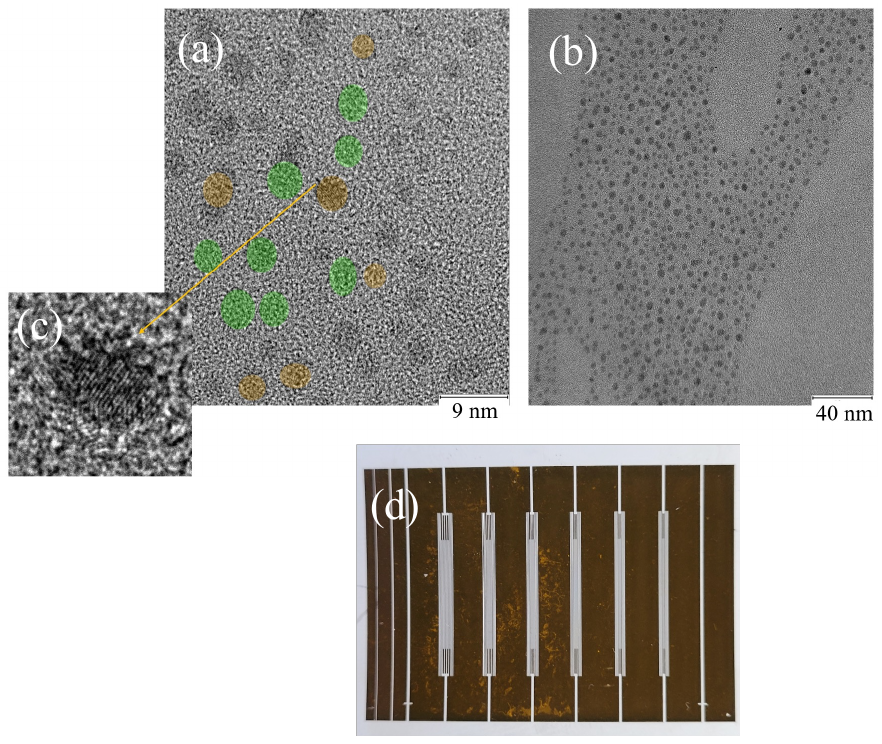}
  \caption{(a) Sample seen under TEM with a typical scale of 9~nm.
  The yellow particles are crystalline $sp^2$ CQDs with a lattice spacing of about $1.9$~\AA.
  The green particles are amorphous $sp^3$ CQDs. (b) TEM picture with
  a larger scale of 40~nm. (c) Zoomed picture of (a).
  (d) Interdigitated electrodes for transport measurements.
  }
  \label{fig_sample}
\end{figure}

Our devices are based on a thin varistor architecture that comprises interdigitated electrodes with a printable epoxy composite material.
A TEM image shows that the polymer matrix contains many zero-dimensional
CQDs, as depicted in Figs.~\ref{fig_sample}(a) and~(b)
\textcolor{black}{[a zoomed image of one CQD is shown in Fig.~\ref{fig_sample}(c)].
Raman data (see Supplementary Material)
show that $sp^2$ ($sp^3$) dots are ordered (disordered). 
Further, the system is doped with pyridines.}
We stress that our design is more compact that generic surface-mounted-device architectures
since it involves suspending a many-CQD system directly on top of an interdigitated electrode structure [see Fig.~\ref{fig_sample}(d)]. This solution could play a valuable role in new age 3D integrated systems replacing the role of diodes and other cumbersome pieces of hardware, allowing for more flattened and compressed designs.
\textcolor{black}{In the Supplementary Material we show an SEM image of two electrodes near the CQDs.}

We apply a source-drain voltage $V$
and measure the current $I$ at room temperature. Our interdigitated
configuration allows for individual control of each microelectrode, offering enhanced versatility in device operation and purpose, since different polymer formulations offer various $I$--$V$ characteristics~\cite{sato2019injection}. This contact scheme is well suited for layers like our CQD polymer that exhibit thickness variations across their dimensions, since the electric measurement inherently averages the device response over the entire layer. Additionally, interdigitated configurations facilitate higher current values for a given $V$, as compared to conventional two-probe measurements, and have negligible impact on the sample and overall device footprint, making it advantageous for robust applications.

Surprisingly, we observe multiple nonlinear $I$--$V$ phenomena in our devices.
The different transport regimes are rather diverse and involve
distinct charge transfer mechanisms, which we now discuss in detail.
However, a word of caution is at this point needed since not all nonlinear
effects are reproducible after subsequent tests.

\begin{figure*}[t]
  \centering
  \includegraphics[width=0.75\textwidth,clip]{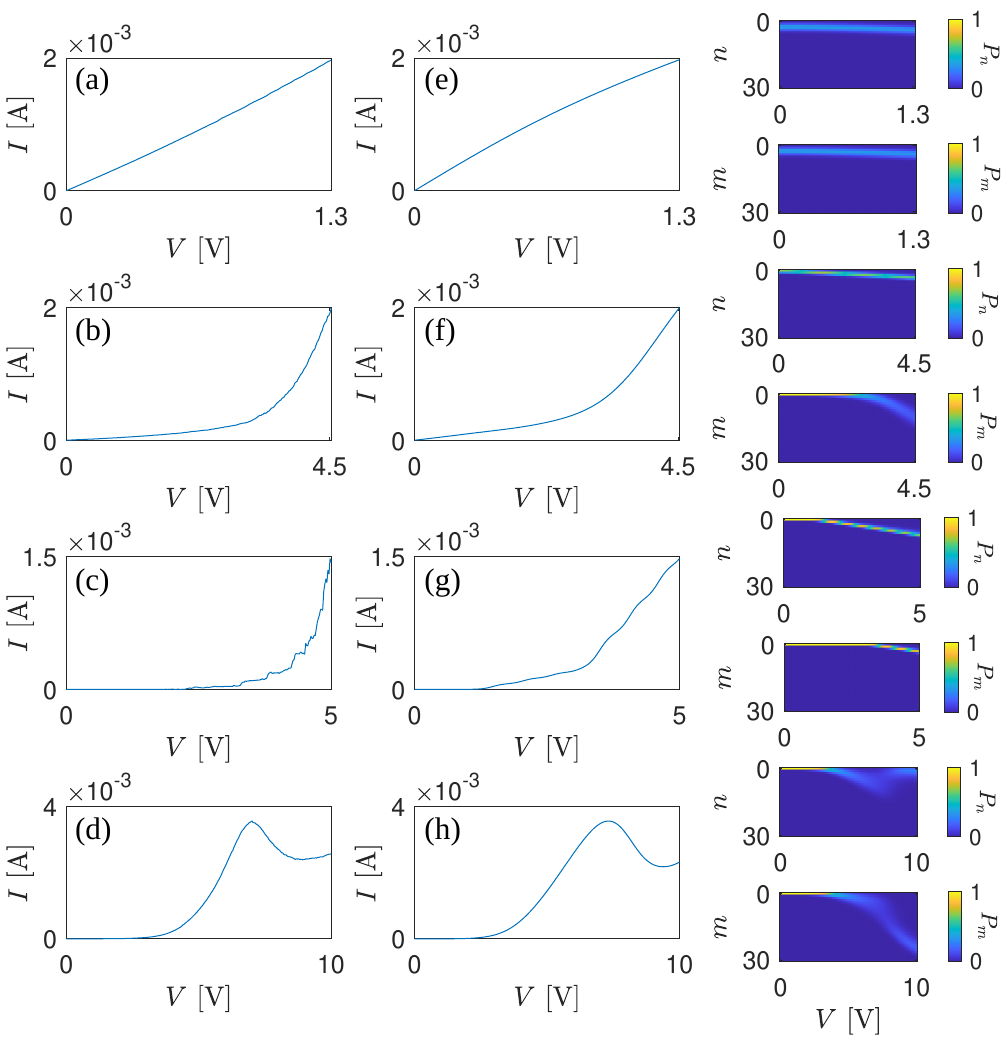}
  \caption{Left panels: experimental results for the current $I$. Middle panels: numerical simulations of the transport model. Right panels: calculated probability $P$ of finding a number
  of excess charges for dot ($n$) and trap ($m$) states. All variables are shown as functions of the applied voltage $V$.
  From top to bottom:
  Ohmic (a,e), diode (b,f), Coulomb blockade (c,g) and negative differential conductance (d,h) transport regimes.
  Parameters for the numerical simulations:  $\Gamma/\gamma=1, \gamma'/\gamma =1, E_d = 1, E_t = 1, C_d =1, C_t=1,C=0, \beta=0.076,I_{\text{max}}=2\times 10^3$ (e),  $\Gamma/\gamma=1, \gamma'/\gamma =1, E_d = 3, E_t = 9, C_d =0.2, C_t=20,C=0, \beta=0.11,I_{\text{max}}=9\times 10^2$ (f), $\Gamma/\gamma=0.2, \gamma'/\gamma =1, E_d = 15, E_t = 35, C_d =0.1, C_t=0.1,C=0, \beta=0.5,I_{\text{max}}=4\times 10^2$ (g), 
  $\Gamma/\gamma=1, \gamma'/\gamma =0.11, E_d = 10, E_t = 11, C_d =1, C_t=1,C=30, \beta=0.1,I_{\text{max}}=3\times 10^3$ (h). Energies are in units of $k_B T$ ($T=290$~K) and capacitances in $e^2/(k_B T)$. $\beta$ is the lever arm factor for voltage and $I_{\text{max}}$ is the current scaling factor. 
  }
  \label{fig_results}
\end{figure*}

The most commonly observed regime is a trivial resistor (Ohmic) behavior, as shown in Fig.~\ref{fig_results}(a). A few samples initially behave as linear resistors on the first test, although this pattern is mostly observed in samples after successive measurements.
In this case, the $\alpha$ coefficient is~1, extracted from the fit
$I=kV^\alpha$ ($k$ is a constant), and is rather low as expected.

Subsequently, we see $I$--$V$ characteristics with higher values
of $\alpha$ [see Fig.~\ref{fig_results}(b) with $\alpha=7$].
This resembles diode behavior and is the most common form of
nonlinearity found while testing. Current sharply increases
for voltages around 3~V. On subsequent tests,
this regime usually appears on the second test, progressively
becoming more monotonic with further measurements.

The third transport regime is illustrated in Fig.~\ref{fig_results}(c).
As can be observed, the $I$--$V$ characteristics
is dominated by steps. This staircase-like behavior
is prototypical of Coulomb blockade~\cite{beenakker1991theory}.
Current jumps are associated to tunneling of
single electrons through the dots
when the tunnel resistance is much larger than $h/e^2$.
Additionally, the electrostatic
energy of an excess electron in a dot
must be larger than temperature.
The fact that our devices exhibit Coulomb blockade
at room temperature implies that the dot
effective capacitance is rather small. We will further
discuss this in the model below. When
the applied voltage surpasses the electrostatic
energy associated to Coulomb repulsion,
a dot can then be charged with an extra electron, the resonance
becomes activated and current flows. The initial average $\alpha=15$
value suggests that Coulomb blockade may play a role in elevating $\alpha$ when the phenomenon is present, but it is worth noting that each step has its own $\alpha$ value, so an average must be derived from every step in the system. 
Analogous blockaded transport has been observed
in organic thin-film transistors~\cite{schoonveld2000coulomb}
but with molecular sites forming a regular lattice whereas
our samples are highly disordered.
This might affect the robustness of this transport
mechanism. The $I$--$V$ steps and elevated $\alpha$ disappear after one or two runs,
presumably due to the high sensitivity of single-particle
transport phenomena to the electron properties
of the tunnel junctions that couple the dots to the contacts.
If the junctions change their morphology from their initial state, Coulomb blockade becomes
hardly reproducible, as already pointed out
in similar nanosized islands~\cite{tedesco2010titanium,caillard2013gold}.
\begin{figure}[t]
  \centering
  \includegraphics[width=0.95\linewidth,clip]{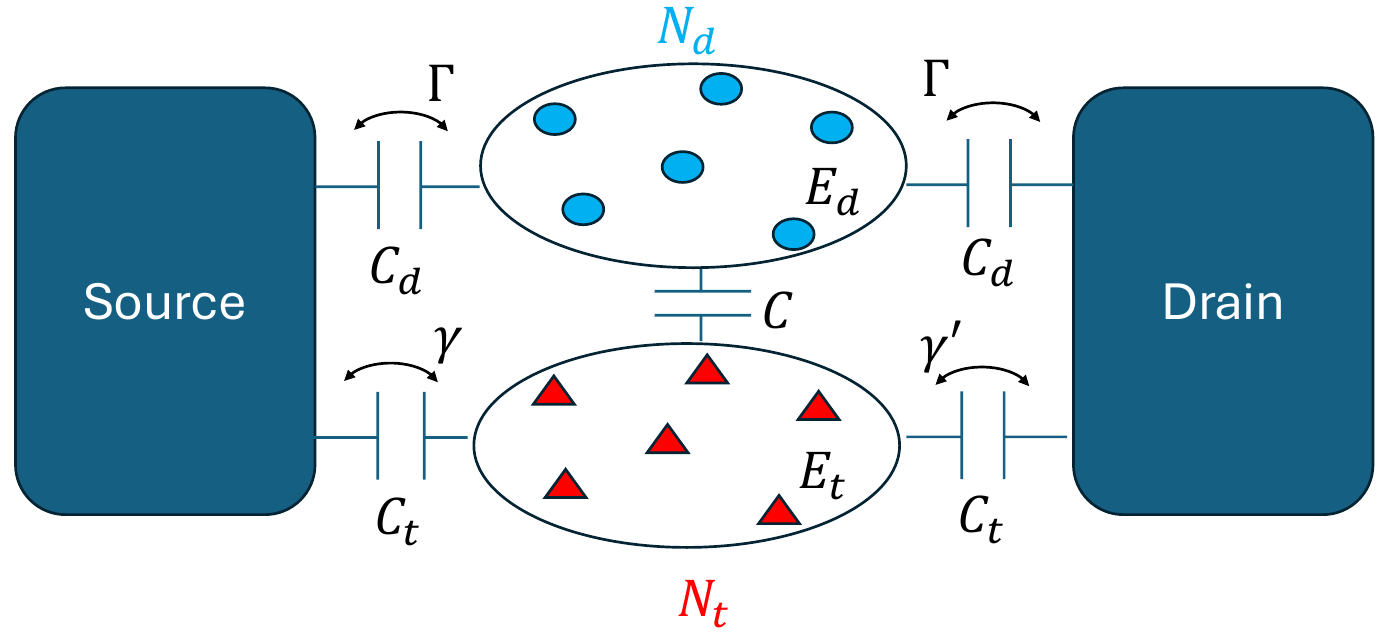}
  \caption{Sketch of the model system. A large number of dot states
  (blue circles) interacts with a large number of trap states (red
  triangles) via the capacitive coupling $C$. In turn, these electronic
  states are both tunnel and capacitively coupled to source
  and drain electrodes via $\Gamma$ or $\gamma$ and $C_d$
  or $C_t$, respectively.
  }
  \label{fig_model}
\end{figure}

Finally, we also observe a negative differential
conductance behavior as displayed in Fig.~\ref{fig_results}(d).
While in the previous three regimes $I$
is always an increasing function of $V$ even if
the $I(V)$ functional dependence is nonlinear
[see Figs.~(b) and~(b)], in this last regime
$I$ grows achieving a maximum at around 6.5~V
(we find in this increasing region $\alpha=7.5$)
but then decreases as voltage further increases. 
This resonance tunneling diode behavior is
typical of planar heterojunctions~\cite{chang1974resonant},
and can lead to exotic nonlinear effects
such as hysteresis~\cite{goldman1987observation} and super-Poissonian noise~\cite{blanter1999transition}.
In single-particle transport, negative differential
conductance has been associated to
multilevel sites with different
coupling strengths to the electrodes~\cite{PhysRevLett.90.076805,thielmann2005super}.
If an electron tunnels into a level (hereafter termed ``trap")
that lies higher than a lower resonance (hereafter termed ``dot")
and the upper level is very weakly coupled to the drain,
the electron becomes trapped and transport through the dot
level cannot occur due to strong repulsion
with the electron in the trap.
\textcolor{black}{Within our system, these traps are induced by the high
degree of structural disorder as shown in Fig.~\ref{fig_sample}(a)
for the CQD distribution.}

We now propose a theoretical framework that covers
the four experimentally found regimes. We illustrate in Fig.~\ref{fig_model}
the system model. The sample contains $N_d$ dot and $N_t$ trap
states. Electron conduction occurs via tunneling across dots and traps,
since the polymer matrix that hosts the CQDs is insulating.
For clarity, we spatially distinguish in the sketch between
dots and traps, although both resonances can be in the same CQD.
Quite generally, their energies ($E_d$ for the dot states
and $E_t$ for the trap states) differ. Electronic repulsion between
both energy levels is modeled with the capacitance $C$.
Charge polarization effects with the contacts are described
with capacitances $C_d$ (dots) and $C_t$ (traps).
Below, we discuss the electrostatic model that describe
the Coulomb blockade phenomenon.
Additionally, tunnel couplings with the source and drain electrodes
are parametrized with widths $\Gamma$ (dot) and $\gamma$ (trap).
These should be understood as statistically averaged couplings.
In reality, due to the large number of states the tunnel widths
will fluctuate according to some distribution (possibly, Porter-Thomas~\cite{jalabert1992statistical}).
For definiteness, we hereafter neglect these coupling fluctuations.
Importantly, traps can couple to the drain more weakly 
than to the source ($\gamma'<\gamma$) to model the trap
mechanism that leads to NDC.

In Coulomb-blockade systems, electrons tunnel sequentially
through the site levels. In other words, quantum coherence
effects can be safely neglected. Therefore, dots and traps
are fully characterized by their occupation numbers
$n$ and $m$, respectively, from which $n+m$
is the total number of excess electrons. The probability distribution
that $n$ dot levels and $m$ trap levels are occupied
evolves over time following the rate equation 
\begin{widetext}
\begin{equation} \label{eq_pnm}
  \begin{aligned}
    \dot{P}_{n,m} =
    & -P_{n,m}
  \Big[ n\Gamma (\tilde{F}^{(S)}_{n,m}+\tilde{F}^{(D)}_{n,m}) 
      +(N_d-n) \Gamma (F^{(S)}_{n+1,m}+F^{(D)}_{n+1,m}) \\
    & \qquad \qquad +m  (\gamma \tilde{f}^{(S)}_{n,m} +\gamma' \tilde{f}^{(D)}_{n,m}) 
      +(N_t-m)  (\gamma f^{(S)}_{n,m+1} +\gamma' f^{(D)}_{n,m+1})  \Big] \\
  & +P_{n+1,m} (n+1) \Gamma (\tilde{F}^{(S)}_{n+1,m}+\tilde{F}^{(D)}_{n+1,m})(1-\delta_{n,N_d}) 
  +P_{n-1,m} (N_d-n+1) \Gamma (F^{(S)}_{n,m}+F^{(D)}_{n,m})(1-\delta_{n,0})\\
  & +P_{n,m+1} (m+1) (\gamma \tilde{f}^{(S)}_{n,m+1}+ \gamma' \tilde{f}^{(D)}_{n,m+1})(1-\delta_{m,N_t})
    +P_{n,m-1} (N_t-m+1) (\gamma f^{(S)}_{n,m}+ \gamma' f^{(D)}_{n,m})(1-\delta_{m,0})
    \end{aligned}
\end{equation}
\end{widetext}
Each term in the right-hand side of Eq.~\eqref{eq_pnm}
describes a charge transfer. For instance, the first term
describes electrons hopping off the dot state through the source ($S$)
or drain ($D$). The rate is proportional to both the number
of filled dots ($n$) and the probability that the contact state is empty,
given by $\tilde{F}=1-F$, with $F$ the Fermi-Dirac distribution function
\begin{align} \label{eq_FS}
    {F}^{(S)} = \frac{1}{1+e^{(\mu_d-eV_S)/k_BT}}\,.
\end{align}
Here, $\mu_d=\mu_d(n,m)$ is the dot electrochemical potential (to be calculated
below), $V_S$ is the source voltage and $T$ is the background temperature,
common to both electrodes. $\tilde{F}^{(D)}$ is obtained
from Eq.~\eqref{eq_FS} by replacing $S$ with $D$.
Analogously, the third term in the right-hand side of Eq.~\eqref{eq_pnm}
describes an electron leaving the trap state, which now depends
on $\tilde{f}=1-f$ with
\begin{align} \label{eq_FStrap}
    {f}^{(S)} = \frac{1}{1+e^{(\mu_t-eV_S)/k_BT}}\,,
\end{align}
where $\mu_t=\mu_t(n,m)$ is the trap electrochemical potential
and the exchange $S\to D$ must be performed when the electron
transfer involves the drain contact. 
The rest of the terms in Eq.~\eqref{eq_pnm}
can be understood in a similar fashion.

Next, we need to calculate an expression for the electrochemical
potentials. These measure the energy required to transfer an additional
electron to the dot or the trap states. To leading order,
repulsion can be treated within the Hartree approximation of
electron-electron interactions, which amounts to describing
interactions with effective capacitance coefficients. We must
consider not only the mutual interaction between dots
and traps given by $C$ but also the energy shifts
due to interactions between the localized levels and
the electrodes, described with $C_d$ and $C_t$
(we permit that dots and traps have different capacitive
couplings since their energies differ). It is important
to take into account the polarization charges induced
by the biased contacts for the theory to be gauge invariant~\cite{buttiker1993capacitance,sanchez2016nonlinear}.
This gives the correct
voltage dependence for the nonlinear conductance
when Coulomb blockade effects are present since current must be a function of $V=V_S-V_D$ only~\cite{stafford1996nonlinear,sanchez2005chirality}. Thus,
following Ref.~\cite{keller2016cotunneling} we find
\begin{align}
    \mu_d&=E_d+\frac{1}{K}\left[\left(
    C_t+\frac{C}{2}    \right) (2n-1) + 2 C m
    \right] \,, \\
    \mu_t&=E_t+\frac{1}{K}\left[\left(
    C_d+\frac{C}{2}    \right) (2m-1) + 2 C n
    \right] \,,
\end{align}
where $K=4 C_d C_t + 2C (C_d+C_t)$. We remark that
the expressions for 
the electrochemical potentials include both a kinetic
term given by the confinement energies and a potential
term that depends on the capacitances and charge configuration.

We focus on the stationary case. Hence, we solve
Eq.~\eqref{eq_pnm} setting $\dot{P}_{n,m}=0$.
The solution is then introduced into the current expression:
\begin{align}\label{eq_I}
    I &= e\sum_{n,m} P_{n,m} \Big[
     n\Gamma \tilde{F}^{(D)}_{n,m}  -(N_d-n) \Gamma F^{(D)}_{n+1,m} \nonumber\\
     &+m  \gamma' \tilde{f}^{(D)}_{n,m} 
      -(N_t-m)  \gamma' f^{(D)}_{n,m+1})  \Big] \,.
\end{align}
Here, we measure the electric current at the drain
electrode. Due to charge conservation, the current
at the source is simply $-I$. There exist
four contributions to the net current as reflected
in the right-hand side of Eq.~\eqref{eq_I}:
a transition $(n,m)\to (n-1,m)$ proportional
to the tunnel rate $\Gamma$ and the number of occupied
dot states $n$ that allows electrons to travel
from the dots to the drain, a transition $(n,m)\to (n+1,m)$ that
contributes with the opposite sign because electrons travel
from the drain to the $N-n$ empty dot states and finally two
further analogous terms for the trap electrons.

The theoretical results
for Eq.~\eqref{eq_I} as a function of voltage
are displayed in Figs.~\ref{fig_results}(e), (f), (g) and (h)
side by side with the experimental $I$--$V$ curves for easy comparison. (For appropriate scaling, the voltage axis
is rescaled by the lever arm $\beta$ whereas the
current axis is multiplied by the factor $I_\text{max}$~\cite{sanchez2007spin}.)
As seen, the agreement is rather good and the various
regimes are obtained by changing the system parameters.
In the rightmost panels, we depict the probability
$P_n=\sum_m P_{nm}$ ($P_m=\sum_n P_{nm}$) that $n$ dots
($m$ traps) are occupied for each of the four $I$--$V$ characteristics.
When solving Eq.~\eqref{eq_pnm} we must
establish a cutoff for $n$ and $m$.
As shown in the color plots, the cutoff
is set at~30. This does not imply that there are only
30~CQDs in our samples. The concentration is indeed much larger.
However, as evidenced by the numerical simulations,
only a few CQD channels effectively participate in the electronic
transport. We recall that Eq.~\eqref{eq_pnm} determines
the excess charges present in the system under the influence
of a voltage bias. The rest of CQDs have background
charges that are confined and therefore do not contribute
to the measured current.

The Ohmic regime [Fig.~\ref{fig_results}(e)] is found when the trap
and dot levels are aligned and both have the same
tunnel couplings (which is equivalent to
having a single type of resonance). Then, the system
response is linear with applied $V$. Interactions play
almost no role because $V$ is much larger than
the charging energies. The number of dots
and traps that become progressively filled
upon application of voltage increases
monotonically, as expected.

The diode behavior [Fig.~\ref{fig_results}(f)] shows up when the two resonances
are clearly separated. At low voltages, transport
is dominated by the lower dot channels. With increasing $V$,
the higher trap channels start to contribute to the electron
flow and current is thus greatly enhanced. We note
the highly asymmetric capacitive couplings. When
the trap states are more strongly coupled to the electrodes,
charge transfer is boosted when $V$ grows.
That this is the mechanism is confirmed by observing
the values of filled $n$ and $m$. The latter increases
dramatically in parallel with the exponential
increase of the current.

The Coulomb-blockade regime [Fig.~\ref{fig_results}(g)]
is dominated by interactions,
as can be seen from the value of the capacitances,
which is ten time smaller than in the Ohmic case
(thereby repulsion is ten times larger).
As a consequence, we obtain a staircase behavior in the $I$--$V$ curves.
The effect is also visible in the $n$ and $m$
panels. The number of filled dot and trap states increases
in steps, a consequence of the discreteness of the electron
charge. 

Finally, the NDC case [Fig.~\ref{fig_results}(h)] is characterized by the interlevel
capacitive coupling, which is nonzero in this regime.
Repulsion between electrons lying in the dot and
trap states combined with asymmetric tunnelings ($\gamma'<\gamma$)
produces the trap mechanism that leads to a current decrease
for a certain voltage interval. For even higher values
of $V$, electrons can be released and current increases again.
This is nicely illustrated in the $n$ panel, where the
number of filled dot states suddenly decreases at around
$V=7$~V and then further increases once more when the applied
voltage is larger than around $V=9$~V. In contrast,
the number of trap states $m$ always increases with $V$.

We have investigated the electron transport
through carbon quantum dots dispersed in a polymer matrix. We have observed
nonlinear $I$--$V$ characteristics in our system and have put forward
a microscopic model that allows us to compute the stationary charges
in the dots when a voltage bias is applied across the system.
This model reinforces our empirical observations by replicating
the diversity of the found transport regimes. Further, our theory
sheds light on the different transport mechanisms involved in
the devices under study. 

Our measurements are restricted to room temperature. Future work
should consider cryogenic temperatures to elucidate the role
of many-body quantum effects beyond Coulomb blockade
(e.g., cotunneling~\cite{de2001electron} or Kondo effect~\cite{goldhaber1998kondo,cronenwett1998tunable,SCHMID1998182})
that could reveal new possibilities in the millikelvin regime.
\textcolor{black}{This could also help mitigate the associated to repeated measurements. Further work should also investigate the ideal distribution of dopants that stabilizes the system, thereby
enabling the acquisition of more reproducible data.}
Another improvement
would be the inclusion of gate electrodes, which offer a more
controllable manipulation of the background charge in both
the dots and the electron traps. Finally,
integration of our systems into current design formats for circuit
boards remains a technological challenge.

\section*{Supplementary material}

\textcolor{black}{See supplementary material for Raman data and SEM contact image.}

\section*{Acknowledgments}

This work was supported by NAMICS, eM-TECH Inc.\
and the Spanish State Research Agency
(MICIU/AEI/10.13039/501100011033) and FEDER (UE) under the Mar{\'\i}a de Maeztu project CEX2021-001164-M.

\section*{Data Availability Statement}

The data that support the findings of
this study are available from the
corresponding author upon reasonable
request.

\bibliography{biblio}

\end{document}